\documentclass[a4paper,12pt]{article}
\usepackage{epsfig}
\title{\Large\bf On Inflation and Variation of
the Strong Coupling Constant}

\author
{ \it \bf  N. Chamoun$^{1,2}$, S. J. Landau$^{3,4}$\thanks{member of
CONICET} , H. Vucetich$^{1,3}$,
\\ \small$^1$ Departamento de F\'{\i}sica, Universidad Nacional de
La Plata,\\ \small c.c. 67, 1900 La Plata, Argentina. \\
\small$^2$ Department of Physics, Higher Institute for Applied
Sciences and Technology,\\ \small P.O. Box 31983, Damascus, Syria.
\\ \small$^3$ Observatorio Astron\'{o}mico, Universidad Nacional
de La Plata,\\ \small Paseo del Bosque S$/$N, CP 1900 La Plata,
Argentina.
\\ \small$^4$ Departamento de F\'{\i}sica, Universidad Nacional de
Buenos Aires,\\ \small  Ciudad Universitaria - Pab. 1, 1428 Buenos Aires,
  Argentina \\
}

\date{}

\begin{document}
\maketitle
\begin{abstract}
Variation of constants in the very early universe can generate
inflation. We consider a scenario where the strong coupling
constant was changing in time and where the gluon condensate
underwent a phase transition ending the inflation.
\end{abstract}



\section{Introduction}
\label{Introduction} Alternatives to inflationary cosmology
\cite{guth81,linde94} include varying speed of light (VSL)
theories \cite{mag03}. Usually all inflationary models are based
on using new fundamental scalar fields, the `inflatons', whose
nature is still unknown. Some models change the matter content of
the universe, while others give the inflaton geometrical
interpretations within brane settings \cite{khoury}. VSL scenarios
may solve the cosmological problems usually tackled by inflation
(``horizon", ``flatness" and ``structure formation" problems)
without introducing inflatons, whereas many inflationary models
lead to variation of other constants of nature \cite{vilenkin99}.
In this paper we will follow the reverse route that variation of
constants in the very early universe can generate inflation. In
an earlier work \cite{chamoun2000}, we considered a Bekenstein-like model for the
QCD strong coupling constant $\alpha_S$ introducing a scalar field
$\epsilon$ expressing the time variation of $\alpha_S$. We found
that experimental constraints going backward till quasar formation
times rule out $\alpha_S$ variability. However, when this model is
implemented in the very early universe, the scalar field
$\epsilon$ can play the role of an inflaton, and one can realize a
consistent inflation scenario  with suitable value for the gluon
condensate. We find that the `time varying' QCD lagrangian leads
naturally to a monomial quadratic potential like the chaotic
scenario. However, while the large values of the inflaton matter
field plague the latter scenario, they just amount in our model to
a reduction of the strong charge by around $10$ times during the
inflation. An exit way can be achieved if the gluon condensate
suffers a phase transition reducing its value to its current value
ending thus the inflation. We will not dwell on the possible
mechanisms for such a phase transition to occur, but wish to
concentrate on the conditions our model should satisfy in order to
present a consistent set up able to accommodate the recent
measurements from the Cosmic Microwave Background (CMB)
\cite{deBernardis,MAXIMA} and the WMAP results \cite{wmap,wmap3}.

\section{Analysis}
\label{analysis} We follow the notations of \cite{chamoun2000}.
Our starting point is the `time varying' QCD Lagrangian
\begin{eqnarray}
L_{QCD}&=& L_{\epsilon}+L_g+L_m \nonumber\\
&=&-\frac{1}{2l^2}\frac{\epsilon_{,\mu}
\epsilon^{,\mu}}{\epsilon^2}
 -\frac{1}{2}Tr(G^{\mu\nu}G_{\mu\nu}) +
\sum_f \bar{\psi}^{(f)}(i \gamma^\mu D_\mu -M_f) \psi^{(f)}
\end{eqnarray}
where $l$ is the Bekenstein scale length, $\epsilon(x)$ is a
scalar gauge-invariant and dimensionless field, with the
`variable' QCD coupling given by $g(x)=g_0\epsilon(x)$. The
covariant derivative is $D_{\mu} =
\partial_{\mu} - i g_0 \epsilon(x) A_{\mu}$ and the gluon tensor field is
$ G^a_{\mu\nu} = \frac{1}{\epsilon} [\partial_\mu(\epsilon
A^a_\nu)-\partial_\nu(\epsilon
A^a_\mu)+g_0\epsilon^2f^{abc}A^b_\mu A^c_\nu] $

Assuming homogeneity and isotropy for an expanding universe we
consider only temporal variations for $\alpha_S\equiv
\frac{g^2(t)}{4\pi}=\alpha_{S_0}\epsilon^2(t)$. One gets the
following equations of motion
\begin{eqnarray}
\label{equation1}
(\frac{G^{\mu\nu}_a}{\epsilon})_{;\mu}-g_0f^{abc}G^{\mu\nu}_b
A^c_\mu+\sum_f g_0\bar{\psi}t^a \gamma^\nu\psi &=&0
\end{eqnarray}
\begin{eqnarray}
\label{equation2} (a^3 \frac{\dot{\epsilon}}{\epsilon})^.
&=&\frac{a^3(t) l^2}{2} \langle G^2 \rangle
\end{eqnarray}
where $a(t)$ is the expansion scale factor in the R-W metric.

Subtracting the total derivative $ \Delta
T^{\alpha\beta}=\partial_{\nu}\left(
-\frac{G_a^{\alpha\nu}}{\epsilon}\epsilon A^{a \beta}\right) $
from the canonical energy-momentum tensor $\frac{\partial
L}{\partial(\partial_{\alpha}\phi_i)}\partial^{\beta}\phi_i -
g^{\alpha\beta}L$ we get the gauge-invariant energy momentum
tensor
\begin{eqnarray}
T^{\alpha\beta}&=& G_a^{\alpha\nu}G_{\nu}^{a\beta}+ i\sum_f
\bar{\psi}^{(f)} \gamma^{(\alpha} D^{\beta
)} \psi^{(f)}
-
\frac{1}{l^2}\frac{\partial^{\alpha}\epsilon
\partial^{\beta}\epsilon}{\epsilon^2} \nonumber\\&&
-g^{\alpha\beta} \left[ -\frac{1}{4}G^{\mu\nu}_a G_{\mu\nu}^a +
\sum_f \bar{\psi}^{(f)}(i \gamma^\mu D_\mu -M_f) \psi^{(f)}
-\frac{1}{2l^2}\frac{\epsilon_{,\mu}\epsilon^{,\mu}}{\epsilon^2}
 \right]
\end{eqnarray}

In the analysis below, we consider only the gauge fields and the
scalar field, i.e. we drop the matter fields in the radiation
dominated very early era. We decompose our energy momentum tensor
into two parts: the gauge part and the $\epsilon$-scalar field
part
\begin{eqnarray}
T_{\alpha\beta}&=&T_{\alpha\beta}^{g}+T_{\alpha\beta}^{\epsilon}
\\&=& \left ( G_{\alpha\nu}G^{\nu}_{\beta}-g_{\alpha\beta}
\left[ -\frac{1}{4}G^{\mu\nu}_a G_{\mu\nu}^a \right] \right) +
\left(
-
\frac{1}{l^2}\frac{\partial_{\alpha}\epsilon
\partial_{\beta}\epsilon}{\epsilon^2}-g_{\alpha\beta} \left[
-\frac{1}{2l^2}\frac{\epsilon_{,\mu}\epsilon^{,\mu}}{\epsilon^2}
\right] \right)\nonumber
\end{eqnarray}
Here all the operators are supposed to be renormalized and it is
essential in the inflationary paradigm that quantum effects are
small in order to get small fluctuations in the CMB.

The contribution of the scalar field to the energy density
$\rho_{\epsilon}=T_{00}^{\epsilon}$ and to the pressure
$T^{\epsilon}_{ij}=g^{(3)}_{ij}p_{\epsilon}$ are
\begin{eqnarray}
\rho_{\epsilon}&=-\frac{1}{2l^2}(\frac{\dot{\epsilon}}{\epsilon})^2=&p_{\epsilon}
\end{eqnarray}

On the other hand, the gauge field contribution
$T_{\alpha\beta}^{g}$ can be decomposed into traceless and trace
parts.

\begin{eqnarray}
\rho_{g} &=& \rho^r_{g} + \rho^{T}_{g} \\
p_{g} &=& p^r_{g} + p^T_{g}
\end{eqnarray}
where $\rho_{g}^{r}, p_{g}^{r}$ are the density and the pressure
corresponding to the ``traceless" part of the gauge field
satisfying $ \rho_{g}^{r} = 3  p_{g}^{r} $, while the trace part
of the gauge field energy-momentum tensor is proportional  to
$g_{\alpha \beta}$ and behaves like a `cosmological constant'
term:
\begin{eqnarray}
\label{impose} \rho^{T}_{g} &=& -p^{T}_{g}
\end{eqnarray}
This equation is reminiscent of `ordinary' inflationary models.
However, to compute the trace part of the density one needs a
`trace anomaly' relation for our `time varying' QCD. Since the
energy-momentum tensor $T^{g}_{\alpha \beta}$ is identical in form
to ``ordinary" QCD and since the trace anomaly which involves only
gauge invariant quantities should, by dimensional analysis, be
proportional to $G^2$, we take it to be the same as in
``ordinary'' QCD (we have checked that changing the numerical
value of proportionality will not alter the conclusions). Thus we
take, up to leading order in the time-varying coupling ``constant"
$\alpha_S=\alpha_{S_0}\epsilon^2$, the relation \cite{greiner95}:
\begin{eqnarray}
\label{traceanomaly}
 T^{\mu g}_\mu &=&
\rho_{g}-3p_{g} = -\frac{9 \alpha_{S_0} \epsilon^2}{8 \pi}
G^{\mu\nu}_a G^a_{\mu\nu}
\end{eqnarray}
This leads to:
\begin{eqnarray}
\label{rhoT} \rho^{T}_{g}&=&-\frac{9 \alpha_{S_0} }{32 \pi}
\epsilon^2 <G^2>
\end{eqnarray}
As we said before, equation (\ref{impose}) suggests, in analogy to
ordinary inflationary models, that the QCD trace anomaly could
generate the inflation. For this, let us assume that the
``trace-anomaly" energy mass density contribution is much larger
than the other densities:
\begin{eqnarray}
\rho_{g}^{T} >> \rho_{\epsilon},\rho^{r}_{g} &\Rightarrow& \rho
\sim \rho_{g}^{T}
\end{eqnarray}
Then, equation (\ref{rhoT}) tells that the vacuum gluon condensate
$<G^2>$ should have a negative value which is not unreasonable
since the inflationary vacuum has ``strange" properties. In
ordinary inflationary models, it is filled with repulsive-gravity
matter turning gravity on its head \cite{guth01}. This reversal of
the vacuum properties is reflected, in our model, by a reversal of
sign for the vacuum gluon condensate.

Now we seek a consistent inflationary solution  to the FRW
equations in a flat space-time:
\begin{eqnarray}
\label{frw}
(\frac{\dot{a}}{a})^2&=& \frac{8 \pi G_N}{3} \rho\\
\frac{\ddot{{a}}}{a}&=&-\frac{4 \pi G_N}{3}(\rho+3 p)
\end{eqnarray}
where $G_N$ is Newton's constant. The first FRW equation with
(\ref{rhoT}) will give

\begin{eqnarray}
\label{H}
H \equiv \frac{\dot{a}}{a} &=& \epsilon\sqrt{\frac{3 \alpha_{S_0} }{4} G_N |<G^2>|}
\end{eqnarray}
On the other hand, the equation of motion (eq.\ref{equation2}) of the scalar field  can be expressed in the following way:
\begin{eqnarray}
\label{epsilon}
\frac{\ddot{\epsilon}}{\epsilon} + 3 H \frac{\dot\epsilon}{\epsilon}
-(\frac{\dot{\epsilon}}{\epsilon})^2 = \frac{l^2 <G^2>}{2}
\end{eqnarray}
This equation differs from the ordinary `matter' inflationary
scenarios in the term $(\frac{\dot{\epsilon}}{\epsilon})^2$.
However, for ``slow roll'' solutions we neglect the terms
involving $\frac{\ddot\epsilon}{\epsilon}$ and
$(\frac{\dot\epsilon}{\epsilon})^2$ to get
\begin{eqnarray}
\label{potential} 3 H \dot\epsilon &=\frac{l^2 <G^2>}{2}
\epsilon&=-V'(\epsilon)
\end{eqnarray}
which is the same as the ``slow roll'' equation of motion of the
inflaton in ordinary scenarios. In our model, the ``slow roll''
condition can be written as:
\begin{eqnarray}
\label{conscond1} \delta \equiv |\frac{\dot{\epsilon}}{H
\epsilon}| &=\frac{2}{9 \alpha_{S_0}}
(\frac{l}{L_P})^2\frac{1}{\epsilon^2}&<< 1
\end{eqnarray}
We set $\epsilon_f$, the value of $\epsilon$-field at the end of
inflation $t_f$, to $1$ so that the time evolution of the strong
coupling terminates with the end of inflation and we expect for
``slow roll'' solutions that $\epsilon_i$, the value of $\epsilon$
at the initial time of inflation $t_i$ corresponding to when the
CMB modes freezed out, to be of order $1$. If the gluon condensate
value $<G^2>$ stays approximately constant during much of the
inflation, the changes of the Hubble constant and the energy mass
density are not very large. In this case we have

\begin{eqnarray}
\label{epsevo} \epsilon (t) & = & \epsilon_i
-\frac{1}{3^{3/2}(\alpha_{S_0})^{1/2}}(\frac{l}{L_p})^2 G^{1/2}_N
|<G^2>|^{1/2} (t-t_i)
\end{eqnarray}
and , as in chaotic scenarios, we get a simple quadratic
potential:
\begin{eqnarray}
V(\epsilon)&=&\frac{l^2 |<G^2>|}{4}\epsilon^2
\end{eqnarray}

One can make explicit the correspondence between our model with
$\epsilon$-scalar field and the chaotic scenario with an
$\phi$-inflaton matter field. Comparing equations (\ref{H}) and
(\ref{potential}) with the corresponding relations in ordinary
inflationary models:
\begin{eqnarray}
 H^2 &=& \frac{8\pi}{3 M_{pl}^2}G_N {\cal V}(\phi) \\
 3 H \dot\phi &=&-{\cal V}'(\phi)
\end{eqnarray}
we find the relations between $(\epsilon,V(\epsilon))$ and
$(\phi,{\cal V}(\phi))$:
\begin{eqnarray}
\phi = \frac{\sqrt{y}}{l}\epsilon &{\rm with}& y=\frac{9 \alpha_{S_0}}{8 \pi} \\
\frac{y}{l^2}V(\epsilon)&={\cal V}(\phi) =& \frac{l^2 |<G^2>|}{4}
\phi^2
\end{eqnarray}

\section{Results and Conclusion}
Now, we check that our model is able to fix the usual problems of
the standard (big bang) cosmology. First, in order to solve the
``horizon" and ``flatness" problems we need an inflation
$\frac{a(t_f)}{a(t_i)}$ of order $10^{28}$ implying an inflation
period $\Delta t = t_f - t_i$ such that
\begin{eqnarray}
\label{hubbletime} H \;\; \Delta t&\sim& 65
\end{eqnarray}
Furthermore, it should satisfy the constraint
\begin{eqnarray}
\label{timecons} 10^{-40} s \leq \Delta t \ll 10^{-10} s
\label{times}
\end{eqnarray}
 so that not to conflict with the explanation of the
baryon number and not to create too large density fluctuations
\cite{hawking82,hawking85}. The bound $10^{-10} s$ corresponds to the
time, after the big bang, when the electroweak symmetry
breaking took place. Presumably, our inflation should have
ended far before this time.
Thus, from equations (\ref{hubbletime}), (\ref{times}) and (\ref{H}) we get the following
bounds on $|<G^2>|$:
\begin{eqnarray}
\label{bound1} 3 \times 10^7 GeV^2 \ll \epsilon |<G^2>|^{1/2} \leq
3 \times 10^{37} GeV^2
\end{eqnarray}
In order to determine $\epsilon_i$, we have $\frac{d \ln a}{d
\epsilon} = \frac{H}{\dot{\epsilon}} \simeq
-\frac{3H^2}{V'(\epsilon)} \simeq -\frac{8 \pi \rho^T_g}{M_{Pl}^2
V'}$ which gives, using equations (\ref{rhoT}) and
(\ref{potential}), the relation:
\begin{eqnarray}
\label{ei} 65&\sim \ln \frac{a(t_f)}{a(t_i)}=&(\frac{L_P}{l})^2
\frac{9 \alpha_{S_0}}{4}(\epsilon_i^2 -1)
\end{eqnarray}
Next, comes the ``formation of structure"
problem and we require the
fractional density fluctuations at the end of
inflation to be of the order $\frac{\delta M}{M}\mid _{t_f} \sim
10^{-5}$ so that quantum fluctuations
in the de Sitter phase of
the inflationary universe form the source of perturbations
providing the seeds for galaxy formation and in order to agree
with the CMB anisotropy limits. Within the relativistic theory of
cosmological perturbations \cite{brandenberger97}, the above
fractional density fluctuations represent (to linear order) a
gauge-invariant quantity and
 satisfy the equation
\begin{eqnarray}
\label{fluctfinal1} \frac{\delta M}{M}\mid _{t_f} &=& \frac{\delta
M}{M}\mid _{t_i}\frac{1}{1+\frac{p}{\rho}}\mid _{t_i}
\end{eqnarray}
where $\delta M$ represent the mass perturbations and where the
initial fluctuations are generated quantum mechanically and are
given by: \cite{brandenberger97,mukhanov92}
\begin{eqnarray}
\label{fluctinitial} \frac{\delta M}{M}\mid _{t_i}= \frac{{\cal
V}'(\Phi)H}{\rho}&=&\sqrt{y}\frac{V'(\epsilon)H}{l\rho}
\end{eqnarray}
whence
\begin{eqnarray}
\label{fluctfinal2} 10^{-5} \sim \mid \frac{\delta M}{M}\mid
_{t_f} &=& \sqrt{y} \mid \frac{V' H}{l} \mid _{t_i} \frac
{1}{|(\rho+p)|_{t_i}}
\end{eqnarray}
In order to evaluate $(\rho+p)|_{t_i}$ we use the energy
conservation equation:
\begin{eqnarray}
\label{enercon} \dot{\rho}+3 (\rho+p) \frac{\dot{a}}{a} &=&0
\end{eqnarray}
and after substituting $\rho \sim \rho^T_{g}$ we get
\begin{eqnarray}
\label{initial}
|(\rho+p)|_{t_i} &=& \frac{1}{24 \pi} (\frac{l}{L_P})^2 |<G^2>|
\end{eqnarray}
In fact, the energy conservation equation can be used to solve for
$\rho^r_{g}$ and we could check that
\begin{eqnarray}
\rho^r_{g} (\dot{\rho^r_{g}}) \sim \rho_{\epsilon}
(\dot{\rho_{\epsilon}}) &\sim& \delta \times \rho ^T_{g} (\delta
\times \dot{\rho^T_{g}})
\end{eqnarray}
where $\delta \equiv |\frac{\dot{\epsilon}}{H\epsilon}| \sim
\frac{1}{\epsilon^2}(\frac{l}{L_P})^2$ and so, when the ``slow
roll'' condition (\ref{conscond1}) is satisfied, our solution
assuming the predominance of the ``trace-anomaly" energy mass
density is self-consistent. Substituting equation (\ref{initial})
in (\ref{fluctfinal2}) and using equation (\ref{potential}) we get
\begin{eqnarray}
\label{important} \frac{l}{L_P} &\sim& 9 \sqrt{\frac{3\pi}{2}}
\alpha_{S_0} 10^{5} \epsilon^2 |<G^2>|^{\frac{1}{2}} G_N
\end{eqnarray}
Hence, taking $G_N \sim 10^{-38} GeV^{-2}$ we obtain
\begin{eqnarray}
\label{G2ei} \frac{l}{L_P} \sim
\frac{|<G^2>|^{\frac{1}{2}}}{10^{34} GeV^2} \epsilon^2
\end{eqnarray}
and combining this last result with (\ref{bound1}), we get
\begin{eqnarray}
10^{-27} &<<& \frac{1}{\epsilon}\frac{l}{L_p} \leq 10^3
\end{eqnarray}
The ``slow roll'' condition (\ref{conscond1}) is consistent with
the upper bound, while the lower bound restricts $\epsilon_i$ not
to be too large.

On the other hand, it is possible to calculate the spectral index of the primordial power
spectrum for a quadratic potential as follows:
\begin{eqnarray}
n-1=-4\eta &{\rm where}& \eta = \frac{M_{Pl}^2}{8\pi}\frac{{\cal
V}''}{{\cal V}}=\frac{2}{9 \alpha_{S_0}}
(\frac{l}{L_P})^2\frac{1}{\epsilon^2}=\delta
\end{eqnarray}
and we find:
\begin{eqnarray} \label{spectral} n &=&
1-\frac{1}{\pi y} (\frac{l}{L_p})^2 \frac{1}{\epsilon_i ^2}
\end{eqnarray}
The inflation would end ($\epsilon_f = 1$) when the ``slow roll''
parameter $\eta = \delta = 1$. We should evaluate the QCD coupling
constant $ \alpha_{S_0}(\mu) = \frac{4
\pi}{\beta_0\ln(\frac{\mu^2}{\Lambda_{QCD}^2})}$ at an energy
scale corresponding to the inflationary period. We take this to be
around the GUT scale $\sim 10^{15}GeV$ and
$\beta_0=11-\frac{2}{3}n_f=7$ (the weak logarithmic dependence
would assure the same order of magnitude for $\alpha_{S_0}$
calculated at other larger scales). With $\Lambda_{QCD}\sim 0.2
GeV$ \cite{data99} we estimate $\alpha_{S_0}\sim 0.025$, and so we
get
\begin{eqnarray} \label{loverL} (\frac{l}{L_P})^2&\sim& 10^{-1}
\end{eqnarray}
This is in disagreement with Bekenstein assumption that $L_P$ is
the shortest length scale in any physical theory. However, it
should be noted that Beckenstein's framework is very similar to
the dilatonic sector of string theory and it has been pointed out
in the context of string theories\cite{strings00} that there is no
need for a universal relation between the Planck and the string
scale. Furthermore, determining the order of magnitude of
$\frac{l}{L_P}$ is interesting in the context of these theories.

From (\ref{ei}), we have $\epsilon_i \sim 11$, and then using
(\ref{spectral}), we have $n=0.97$ which is within the range of
WMAP results \cite{wmap,wmap3}.

The model reproduces the results of the chaotic inflationary
scenario. However, the shape of the potential was not put by hand,
rather a gauge theory with a changing coupling constant led
naturally to it. Moreover, in typical chaotic models, the inflaton
field starts from very large values ($\phi_i \sim 15 M_{Pl}$) and
ends at around $1 M_{Pl}$. One might suspect whether field theory
is reliable at such high energies. Nonetheless, this problem is
absent in our model since the large values have another meaning in
that they just refer to a reduction of the strong coupling by
around 10 times during the inflation.

Furthermore, chaotic inflations get a typical reheating of order
$T_{rh} \sim 10^{15} GeV$, and one might need to worry about the
relic problem. Similarly, equation (\ref{G2ei}) leads in our model
to a gluon condensate $|<G^2>|_i \sim 10^{62} GeV^4$ at the start
of inflation. From equation (\ref{epsevo}), we see that this
corresponds to an inflation time interval $\Delta t \sim
10^{-35}s$ satisfying the constraint (\ref{timecons}). If the
gluon condensate stays constant, as we assumed in our analysis, we
will have the same reheating temperature as in chaotic models
($T_{rh} \sim \rho(t_f)^{1/4}$). However, we should compare this
value for $<G^2>$ with its present value renormalized at GUT scale
$\sim 10^{15} GeV$ which can be calculated knowing its value at
$1GeV$ \cite{chamoun2000} and that the anomalous dimension of
$\alpha_S G^2$ is identical ly zero. We get
\begin{eqnarray}\label{condNow}<G^2(now, \mu\sim 10^{15} GeV)>\sim 1 GeV^4 \end{eqnarray} which
represents a decrease of $62$ orders of magnitude.

\begin{figure}[ht]
\begin{center}
\begin{tabular}{cc}
\epsfig{file=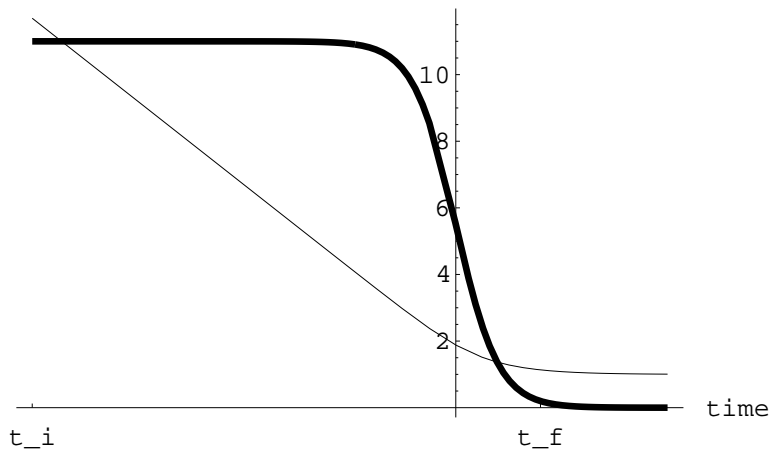}
\end{tabular}
\end{center}
\caption{Temporal evolution of the condensate $|<G^2>|$ (thick
line) and the $\epsilon$-field (thin line), for the choice $f(t) =
-\beta^2 \tanh ^2(\epsilon -1)$. The $<G^2>$ scale has been
adapted so that to visualize both graphs together.} \label{figure}
\end{figure}

This can give a  possible picture for an exit scenario. Lacking a
clear theory for the non-perturbative dynamics of the gluon
condensate, we consider its value $|<G^2>|$ depending on energy,
and thus implicitely on cosmological time, as given by the
standard RGE which turns it off logarithmically at high energy.
However, we can furthermore assume the condensate value  to depend
explicitely on time during inflation:
\[ <G^2(E,t)> = <G_0^2(E(t))> f(t)
\]
where $<G_0^2(E)>$ is the piece determined by the RGE, the unknown
function $f(t)$ should be such that it varies slowly during most
of the inflationary era, to conform with an approximately constant
huge and negative value of $<G^2>$, while at the end of inflation
it causes a drastic drop of the condensate value $<G^2>$ to around
zero. The energy release of this helps in reheating the universe,
while reaching the value $0$ leads to a minute ``trace-anomaly"
energy mass density (equation \ref{rhoT}) ending, thus, the
inflation. The other types of energy density would contribute to
give the gluon condensate its `small' positive value of
(\ref{condNow}), and the subsequent evolution is just the standard
one given by RGE. Surely, this phenomenological description needs
to be tested and expanded into a theory where the concept of
symmetry breaking of such a phase transition for the condensate
$<G^2>$ provides the physical basis for ending the inflation.
Nonetheless, with a test function of the form $f(t) = -\beta^2
\tanh ^2(\epsilon -1)$ with $\beta \sim 10^{31}$, one can
integrate analytically the equation of motion, and in ``slow
roll'' regime we have $\epsilon = 1 + Arcsinh (\exp [- \alpha
\beta t])$ with $\alpha \beta \sim 10^{12} GeV$. The graph in Fig.
\ref{figure} shows the time evolutions of the condensate and the
$\epsilon$-field, which agree with the required features. This
example is meant to be just a proof of existence of such
functions, and the temporal dependence of the condensate could be
of complete different shape while the whole picture is still
self-consistent. The issue demands a detailed study for the
condensate within an underlying theory and we do not further it
here. We hope this work will stimulate interest in the subject.

\section*{Acknowledgements}
This work was supported in part by CONICET, Argentina. N. C.
recognizes economic support from TWAS.

\end{document}